\documentclass{article}
\usepackage{amssymb}
\usepackage{utopia}
\usepackage{amsmath}
\begin{document}

\newcommand{\shalf}{\hbox{${\textstyle{\frac{1}{2}}}$}}
\newcommand{\sthird}{\hbox{${\textstyle{\frac{1}{3}}}$}}
\newcommand{\stwothird}{\hbox{${\textstyle{\frac{2}{3}}}$}}
\newcommand{\sthreequarter}{\hbox{${\textstyle{\frac{3}{4}}}$}}
\newcommand{\trans}{\hbox{${\scriptscriptstyle\top}$}}
\newcommand{\squarter}{\hbox{${\textstyle{\frac{1}{4}}}$}}

\title{Classical Analysis of the van Dam -- Veltman Discontinuity}
\author{Matteo Carrera and Domenico Giulini    \\
   \mbox{}                                     \\
\normalsize{Fakult\"at f\"ur Physik}           \\
\normalsize{Universit\"at Freiburg}            \\
\normalsize{Hermann-Herder Strasse 3}          \\
\normalsize{D-79104 Freiburg, Germany}         \\
   \mbox{}                                     \\
{\small carrera@physik.uni-freiburg.de}        \\
{\small giulini@physik.uni-freiburg.de}
}
\date{\normalsize{July 16th 2001}}
\maketitle

\begin{abstract}
\noindent
We consider the classical theory of a gravitational field with spin 2 and
(Pauli-Fierz) mass $m$ in flat spacetime, coupled to electromagnetism and
point particles. We establish the law of light propagation and calculate
the amount of deflection in the background of a spherically symmetric
gravitational field. As $m$ tends to zero, the deflection is shown
to converge to 3/4 of the value predicted by the massless theory
(linearized General Relativity), even though the spherically
symmetric solution of the gravitational field equations has no regular
limit. This confirms an old argument of van Dam and Veltman on a purely
classical level, but also shows its subtle nature.
\end{abstract}

\subsection*{Notation and Conventions}
We work in flat Minkowski space with metric $\eta=\mbox{diag}(1,-1,-1,-1)$.
Spacetime indices, running from $0$ to $3$ are greek, whereas space
indices, running from $1$ to $3$ are latin. The Minkowski inner product
is denoted by $\eta(p,q)=p\cdot q=p^{\mu}q_{\mu}$, and $p\cdot p=p^2$.
Partial differentiation with respect to $x^{\mu}$ is either written as
$\partial_{\mu}$ or sometimes simply as $,\mu$. Throughout we use units
in which the velocity of light is set to 1.

%-----------------------------------------------------------------------

\section*{Introduction}
According to an old and well known argument of van Dam and
Veltman~\cite{vDV1,vDV2}, which we will sketch below, the theory of
massive spin 2 fields in flat Minkowski spacetime does not approximate
the strictly massless theory (linearized General Relativity (GR)),
in the limit as the mass tends to zero. (See also \cite{vD} and the more
comprehensive account in \cite{BD}.) This means that there exist
corresponding observables in both theories which distinguish the massless
limit of the massive theory from linearized GR. One such observable,
so it is claimed, is the amount of deflection of a light ray in the
background of a, say, rotationally symmetric, static gravitational
field. More precisely, in the limit the mass tends to zero,
the deflection angle predicted by the massive theory tends to 3/4 of the
angle predicted by linearized GR.

As we shall see below, the argument given by van Dam and Veltman is
entirely based on the properties of free propagators, whose structure is
determined by Poincar\'e invariance. The aim in this paper is to
improve our understanding of the classical aspects of this limit,
which is known to be non trivial for several reasons,
but has not been explored in all details in the mentioned
references. In particular, we wish to understand how certain
observables can have a smooth limit as the mass tends to zero, even
though they refer to solutions which diverge in that limit.
The deflection of light is an example of such a case, whose
derivation from first principles we shall consider in this paper.

Let us now turn to the argument proper of van Dam and Veltman:
In momentum space, the free propagators are given by
\begin{eqnarray}
P^m_{\mu\nu\alpha\beta}(p) &=&
\frac{\shalf(\pi_{\mu\alpha}\pi_{\nu\beta}  +
             \pi_{\mu\beta} \pi_{\nu\alpha}) -
      \sthird\pi_{\mu\nu}   \pi_{\alpha\beta}}
{p^2-m^2+i\varepsilon},\qquad \mbox{massive case}
\label{eq:massive-propagator}
\\
P_{\mu\nu\alpha\beta}(p) &=&
\frac{\shalf(\pi_{\mu\alpha}\pi_{\nu\beta}  +
             \pi_{\mu\beta} \pi_{\nu\alpha}) -
      \shalf\pi_{\mu\nu}   \pi_{\alpha\beta}}
{p^2+i\varepsilon},\qquad \mbox{mass-less case}
\label{eq:massless-propagator}
\end{eqnarray}
where $\pi_{\mu\nu}$ is the induced Riemannian metric on
the mass-hyperboloid $p^2=m^2$ in momentum space:
\begin{equation}
\pi_{\mu\nu}(p):=\frac{p_{\mu}p_{\nu}}{m^2}-\eta_{\mu\nu}\,.
\label{eq:mass-shell-projector}
\end{equation}
We now consider two systems with (Fourier transformed) energy-momentum
tensors $T^{\mu\nu}$ and $T'_{\mu\nu}$. If we assume these tensors
to be conserved, $0=p^{\mu}T_{\mu\nu}(p)=p^{\mu}T_{\mu\nu}'(p)$, the
one-graviton interaction takes the form (we write $T=T^{\mu}_{\mu}$ etc.)
\begin{equation}
\kappa\,
T^{\mu\nu}(p)P^m_{\mu\nu\alpha\beta}(p)\,{T'}^{\alpha\beta}(-p)=
\kappa\,\frac{T^{\mu\nu}(p)
T_{\mu\nu}'(-p)-\sthird T(p){T'}(-p)}{p^2-m^2+i\varepsilon}
\label{eq:massive-interaction}
\end{equation}
in the massive case, and
\begin{equation}
\kappa_0\,
T^{\mu\nu}(p)P_{\mu\nu\alpha\beta}(p)\,{T'}^{\alpha\beta}(-p)=
\kappa_0\,\frac{T^{\mu\nu}(p)
T_{\mu\nu}'(-p)-\shalf T(p){T'}(-p)}{p^2+i\varepsilon}
\label{eq:massless-interaction}
\end{equation}
in the massless case, with gravitational constants $\kappa$ and
$\kappa_0$ respectively. For slowly-moving massive objects the leading
component of the energy-momentum tensor is $T_{00}$ and we get
$\stwothird \kappa T^{00}(p)T'_{00}(-p)$ in the massive
and $\shalf \kappa_0 T^{00}(p)T'_{00}(-p)$ in the massless case.
To identify $\kappa$ and $\kappa_0$ one requires that both cases
lead to Newton's law of attraction with the \emph{same} Newtonian
constant $G$ (taking the limit $m\rightarrow 0$ in the massive case).
This gives
\begin{equation}
\kappa=\sqrt{\sthreequarter} \kappa_0=\sqrt{12\pi G}.
\label{eq:gravitational-konstant}
\end{equation}

On the other hand, if we consider the interaction of light and slowly
moving matter, there are no trace-terms in (\ref{eq:massive-interaction})
and (\ref{eq:massless-interaction}) due to the tracelessness of the
electromagnetic energy momentum tensor.
But now (\ref{eq:gravitational-konstant}) implies that the massive theory,
in the limit $m\rightarrow 0$, leads to an interaction of light and matter
which is weaker by a factor $3/4$ as compared to the massless theory.
Accordingly, in that limit, light deflection comes out smaller by the
factor $3/4$ as compared to the massless theory, i.e., linearized
General Relativity. This is often taken as sufficient justification
to rigorously argue that present-day observations exclude a finite
graviton mass.

Here one should stress that the flatness of spacetime enters
in an essential way. In fact, it was recently argued that there is
no discontinuous behaviour at $m=0$ if the cosmological constant
$\Lambda$  is non-zero. For negative $\Lambda$ (anti de~Sitter
spacetime) a smooth limit of propagator residues was shown in~\cite{P}.
For positive $\Lambda$ (de~Sitter spacetime) the situation is again
different, since there is no quantum theory of spin 2 fields in the
mass range $0<m^2<\stwothird\Lambda$ obeying unitarity and certain
locality requirements~\cite{H1,H2}. Here a discontinuity occurs
at $m^2=\stwothird\Lambda$.

%--------------------------------------------------------------------

\section{Massive spin-2 fields in Minkowski space}
We briefly review the theory of massive spin-2 fields
in flat spacetime. It is represented by a symmetric tensor field
$h_{\mu\nu}$. The space of such fields still represents the Poincar\'e
group in a highly reducible fashion. A projection into an irreducible
subspace of pure spin-2 and mass $m$ is given by
\begin{eqnarray}
    (\square+m^2)h_{\mu\nu} &=& 0 \label{eq:Klein-Gordon}  \\
\partial^{\mu}h_{\mu\nu} &=& 0 \label{eq:divergenceless}\\
           h^{\mu}_{\mu} &=& 0 \label{eq:tracelessness}
\end{eqnarray}
%

%---------------------------------------------------------------

\subsection{Static rotationally symmetric solution}
We seek a solution of
(\ref{eq:Klein-Gordon},\ref{eq:divergenceless},\ref{eq:tracelessness})
which is static, i.e. $\partial_0h_{\mu\nu}=0$ and rotationally
symmetric in the following sense:
\begin{equation}
D(R)^{\alpha}_{\mu}D(R)^{\beta}_{\nu}\,h_{\alpha\beta}
(D(R)^{\lambda}_{\sigma}x^{\sigma})=h_{\mu\nu}(x^{\lambda})
\quad\forall R\in SO(3)\,,
\label{eq:rotation-invariance}
\end{equation}
where, in a 1+ \!3 - split matrix notation,
\begin{equation}
D(R)=
\left(
\begin{array}{cc}
1 & {\vec 0}^{\trans}\\
\vec 0 & R
\end{array}
\right)\,.
\label{eq:SO(3)-implementation}
\end{equation}
For physical reasons (finite energy around spatial infinity) we also
require the asymptotic fall-off condition
$\lim_{r\to \infty} h_{\mu\nu}(x)=0$, where $r$ is the 3-dimensional
radius. The $00$-component of (\ref{eq:Klein-Gordon}) has the
one-parameter ($b$) family of solutions
\begin{equation}
h_{00}(\vec{x}) = -bf(r) := -b\,\frac{\exp(-mr)}{r},
\label{eq:00-solution}
\end{equation}
where the minus sign is introduced for later
convenience. The $0$-component of (\ref{eq:divergenceless}) and
(\ref{eq:rotation-invariance})
imply $h_{0i}=ax^i/r^3$, which contradicts (\ref{eq:Klein-Gordon})
unless $a=0$; hence $h_{0i}=0$. To determine the spatial components,
we first remark that any rotationally symmetric two-tensor in space is
of the form
\begin{equation}
h_{ij}({\vec x})=f_1(r)\delta_{ij}+f_2(r)\frac{x^ix^j}{r^2}\,.
\label{eq:general-form}
\end{equation}
Equation (\ref{eq:Klein-Gordon}) now reduces to two coupled ODEs
for $f_1$ and $f_2$, which may be decoupled by introducing the new
function ${\tilde f}_1:=3f_1+f_2$. One obtains
\begin{eqnarray}
&& (\Delta-m^2) {\tilde f}_1=0\,,
\\
&& (\Delta-m^2-\tfrac{6}{r^2})f_2=0\,,
\label{eq:dec-ode}
\end{eqnarray}
which, under the given fall-off conditions, have the unique
2-parameter set of solutions
\begin{eqnarray}
{\tilde f}_1(r) &=& c_1\,\frac{\exp(-mr)}{r}\,,
\label{eq:f-1-solution}
\\
f_2(r) &=& c_2\,\frac{\exp(-mr)}{r}\left(1+\frac{3}{mr}+\frac{3}{(mr)^2}\right)\,.
\label{eq:f-2-solution}
\end{eqnarray}
Condition (\ref{eq:divergenceless}) and (\ref{eq:tracelessness})
now imply $c_2=-\shalf c_1$ and $c_1=-b$ respectively, thereby
projecting out a unique one-parameter set of solutions, which,
in terms of the function $f$ defined in (\ref{eq:00-solution}),
can be written in the simple form
\begin{equation}
h_{\mu\nu}(\vec x) = -b\,
\left(
\begin{array}{cc}
f(r)   &  {\vec 0}^{\trans}\\
\vec 0 &  \shalf(\delta_{ij}f(r)-\tfrac{1}{m^2}\partial_i\partial_jf(r))\,
\end{array}
\right) .
\label{eq:solution1}
\end{equation}
We note that the $\partial_i\partial_jf$ - part is of a form that would
be pure gauge in the massless theory. However, there is no gauge freedom
in the massive theory and the prefactor, $m^{-2}$, causes this term to
diverge as $m\rightarrow 0$.

A slightly more geometric way to write the solution (\ref{eq:solution1})
is as follows: Let $\vec n=\vec x/r$ be the radial unit vector; we
define the spatial projection tensors $\rho_{ij}:=n_i n_j$ in
radial direction and $\tau_{ij}:=\delta_{ij}-\rho_{ij}$ in the
orthogonal direction, tangential to the spheres of constant $r$.
In terms of these, the spatial part of (\ref{eq:solution1})
takes the form
\begin{equation}
h_{ij}(\vec x) = -\frac{bf(r)}{2}\left[\tau_{ij}+
\frac{1+mr}{(mr)^2}(\tau_{ij}-2\rho_{ij})\right]\,,
\label{eq:solution2}
\end{equation}
which clearly separates the trace part $\propto\tau$, which stays
finite for $m\rightarrow 0$, and the trace free part
$\propto\tau-2\rho$, which diverges as $m$ tends to zero.

%-------------------------------------------------------------

\subsection{Lagrangian formulation and matter couplings}
The previous equations
(\ref{eq:Klein-Gordon},\ref{eq:divergenceless},\ref{eq:tracelessness})
are equivalent to the Euler-Lagrange-equations of the following
action:
\begin{equation}
\begin{split}
S_g=\squarter \int d^4x\,\Bigl[&
h_{\mu\nu,\lambda}h^{\mu\nu,\lambda}
- 2h^{\mu\lambda}_{\quad,\lambda}h_{\mu\nu}^{\quad,\nu}
+ 2h^{\mu\nu}_{\quad,\nu}h_{,\mu}
- h_{,\mu}h^{,\mu} \\
& \quad - m^2(h_{\mu\nu}h^{\mu\nu}-h^2)\Bigr]\,.
\label{eq:action-gravity}
\end{split}
\end{equation}
The mass term is sometimes called the Pauli-Fierz term.
A straightforward calculation of the variational derivative
$E^{\mu\nu}:=\delta S_g/\delta h_{\mu\nu}$ (keeping in mind the
symmetry $h_{\mu\nu}=h_{\nu\mu}$) shows that the conditions
$E^{\mu\nu}=0$, $\partial_{\mu}E^{\mu\nu}=0$, and $E^{\mu}_{\mu}=0$
indeed imply
(\ref{eq:Klein-Gordon},\ref{eq:divergenceless},\ref{eq:tracelessness})
and vice versa, given that $m\not =0$.

The coupling to some specific form of matter is described
by an interaction term
\begin{equation}
S^{\rm int}=-\frac{\kappa}{2} \int d^4x\,h_{\mu\nu}T^{\mu\nu}\,,
\label{eq:matter-coupling}
\end{equation}
where $T^{\mu\nu}$ is the energy-momentum tensor of the matter.
The possibility of an $h T$-coupling is excluded by an argument given
in section (1.4).
Taking the divergence of $\delta(S_g+S_s^{\rm int})/\delta h_{\mu\nu}$
we get
\begin{equation}
m^2\partial_{\mu}(h^{\mu\nu}-\eta^{\mu\nu}h)
=-\kappa\partial_{\mu}T^{\mu\nu}\,.
\label{eq:divergence-cond}
\end{equation}
Let us assume energy-momentum conservation for the matter
field
\begin{equation}
\partial_{\mu}T^{\mu\nu}=0\,.
\label{eq:matter-T-tracelessness}
\end{equation}
Then $\partial_{\mu}h^{\mu\nu}=h^{,\nu}$, which, upon insertion into the
Euler-Lagrange equations and taking the trace, implies that the traces of
$h^{\mu\nu}$ and $T_{\mu\nu}$ must be proportional
\begin{equation}
h=\frac{\kappa}{3m^2}T\,.
\label{eq:trace-proportionality}
\end{equation}
Using (\ref{eq:divergence-cond}) and
(\ref{eq:trace-proportionality}) in the Euler-Lagrange
equations they read
\begin{equation}
-(\square+m^2)h_{\mu\nu}=
\kappa
\left(T_{\mu\nu}-\sthird\eta_{\mu\nu}T\right)
-\frac{\kappa}{3m^2}\partial_{\mu}\partial_{\nu}T
\label{eq:Klein-Gordon-Matter}\,.
\end{equation}
This leads indeed to the propagator (\ref{eq:massive-propagator})
and establishes the background for the arguments given in the
introduction.

%---------------------------------------------------------------

\subsection{Coupling to a point particle}
The action of a free point particle of mass $m_0$ in Minkowski
space is given by $-m_0$ times its arc-length. Choosing any
parameter $\lambda$ to parametrize its world-line,
$z^{\mu}(\lambda)$, we have
\begin{equation}
S_{\rm p} = -m_0\int d\lambda\ \sqrt{\eta_{\mu\nu}\,
\frac{dz^{\mu}(\lambda)}{d\lambda}
\frac{dz^{\nu}(\lambda)}{d\lambda}}\,.
\label{eq:action-particle}
\end{equation}
Its energy-momentum tensor is
\begin{equation}
T_{\rm p}^{\mu\nu}(x) = m_0\int d\tau\ \delta^{(4)}\left(x-z(\tau)\right)
\frac{dz^{\mu}(\tau)}{d\tau}
\frac{dz^{\nu}(\tau)}{d\tau}\,,
\label{eq:em-tensor-particle}
\end{equation}
where the parameter $\tau$ is the arc-length with respect to
the Minkowski metric $\eta$. Since this expression is not
reparametrisation invariant, we are not free to specify it
otherwise. However, we can play the following trick, which turns
out to be of central importance: Consider a new metric on Minkowski
space, given by
\begin{equation}
g_{\mu\nu}=\eta_{\mu\nu}+\kappa h_{\mu\nu}.
\label{eq:physical-metric}
\end{equation}
The arc length w.r.t.\! $g$ is called $s$.
We consistently neglect terms of higher than linear order in
$\kappa$ and may therefore replace $\tau$ by $s$ in
(\ref{eq:em-tensor-particle}), since this term already gets multiplied
by $\kappa$ in (\ref{eq:matter-coupling}). In (\ref{eq:action-particle})
we choose $\lambda=s$, replace $\eta$ by $g-\kappa h$, and expand to
linear order on $\kappa$. The $\kappa$-term then just cancels the
interaction term (\ref{eq:matter-coupling}) and we get
\begin{equation}
S_{\rm p}+S_{\rm p}^{\rm int} = -m_0\int ds\
\sqrt{g_{\mu\nu}(z(s))\frac{dz^{\mu}(s)}{ds}\frac{dz^{\nu}(s)}{ds}}
= -m_0\int ds\,,
\label{eq:total-action-particle}
\end{equation}
which says that the gravitationally interacting particle moves on
geodesics in the metric $g$. \\

In order to determine the still unknown constants $\kappa$ and $b$
in $g_{\mu\nu}$ (see (\ref{eq:solution1}))
we compare now this equation of motion with the Newton equation.
To first order in $\kappa$ this reads, writing a dot for differentiation
with respect to $s$:
\begin{equation}
{\ddot z}^{\mu}=\tfrac{\kappa}{2}\eta^{\mu\nu}
\left(
\partial_{\nu}h_{\alpha\beta}-
\partial_{\alpha}h_{\nu\beta}-
\partial_{\beta}h_{\nu\alpha}
\right)
{\dot z}^{\alpha}{\dot z}^{\beta}\,.
\label{eq:geodesic-in-g}
\end{equation}
In a Newtonian approximation, where we neglect terms of second
and higher powers in the spatial velocities $dz^k/ds$,
this reduces to
\begin{eqnarray}
{\ddot z}^0\!\!\!&=&\!\!\!-\kappa\, h_{00,k}\, {\dot z}^0 {\dot z}^k\,, \\
{\ddot z}^k\!\!\!&=&\!\!\!-\tfrac{\kappa}{2}\, h_{00,k}\, ({\dot z}^0)^2\,.
\label{eq:nonrel-eq-motion}
\end{eqnarray}
Introducing the Minkowski time coordinate $t:= z^0$ we can rewrite
the second equation, using the first one and again neglecting quadratic
terms in the spatial velocities, as :
\begin{equation}
\frac{d^2 \vec z}{dt^2} = - \tfrac{\kappa}{2}\vec\nabla h_{00}
\label{eq:newtonian-motion}\, .
\end{equation}
Thus, recalling the fall-off condition for $h_{\mu\nu}$, we get the
identification
\begin{equation}
\tfrac{\kappa}{2} h_{00}=\Phi ,
\label{eq:newtonian-potential}
\end{equation}
where $\Phi$ is the Newtonian potential. Applied to our solution
(\ref{eq:00-solution}) we can now determine the product of
the constants $\kappa$ and $b$ to be
\begin{equation}
\kappa b = 2 G M ,
\label{eq:kappa-b-G-M}
\end{equation}
where $G$ is the Newtonian gravitational constant and $M$ is the
central mass, i.e., the source of the gravitational field.
With this identification our metric coefficients $g_{\mu\nu}$ are
now entirely determined.

At this point we may also determine the coupling $\kappa$ separately.
For this we return to the field equation (\ref{eq:Klein-Gordon-Matter}),
whose $00$-component for a static source and in the limit
$m\rightarrow 0$ reduces to
\begin{equation}
\Delta h_{00}=\stwothird\kappa\, T_{00}\,.
\label{eq:newtonian-eq}
\end{equation}
In view of (\ref{eq:newtonian-potential}), a comparison with the
Newtonian equation $\Delta\Phi=4\pi G\rho$ (here $\rho=T_{00}$)
leads to
\begin{equation}
\kappa=\sqrt{12\pi G}\,.
\label{eq:kappa-G}
\end{equation}

Finally, we also remark on the fact that instead of
(\ref{eq:action-particle}) one often finds the following
alternative expression for the action of a free particle
(e.g. in \cite{Th}):
\begin{equation}
S_{\rm p} = -\shalf m_0\int d\lambda\ \eta_{\mu\nu}\,
\frac{dz^{\mu}(\lambda)}{d\lambda}
\frac{dz^{\nu}(\lambda)}{d\lambda}\,.
\label{eq:action-particle-alt}
\end{equation}
Like (\ref{eq:action-particle}) it also leads to straight lines in
Minkowski space, but in addition also to the condition that $\lambda$
is (a constant multiple of) $\tau$, the arc length measured with $\eta$.
Since one is only interested in the spacetime paths and not in their
parametrisation, one may use either (\ref{eq:action-particle}) or
(\ref{eq:action-particle-alt}) to determine the free paths.
However, we wish to point out that in the presence of an interaction of
the form (\ref{eq:matter-coupling},\ref{eq:em-tensor-particle})
the latter choice becomes inconsistent. The reason being the following:
Adding the ``free action'' (\ref{eq:action-particle-alt}) with
$\lambda=\tau$ to the interaction (\ref{eq:matter-coupling})
with the energy momentum  tensor being given by
(\ref{eq:em-tensor-particle}), we obtain an expression like
(\ref{eq:action-particle-alt}), where $\eta$ is replaced
by $g$ and the parameter is still $\tau$. But as a result of the fixed
 parametrisation the Euler-Lagrange equations now imply
$h_{\mu\nu}\frac{dz^{\mu}(\tau)}{d\tau}\frac{dz^{\nu}(\tau)}{d\tau}=0$,
which in general will have no solutions. Hence we maintain that the
action for the particle should be written in the square-root
form~(\ref{eq:action-particle}).

%---------------------------------------------------------------

\subsection{Coupling to the electromagnetic field}
The action for the free electromagnetic field in Minkowski space is
given by
\begin{equation}
S_{\rm em}=-\squarter\int d^4x\,
\eta^{\alpha\mu}\eta^{\beta\nu}F_{\alpha\beta}F_{\mu\nu}\,,
\label{eq:action-em}
\end{equation}
where $F_{\mu\nu}=\partial_{\mu}A_{\nu}-\partial_{\nu}A_{\mu}$.
The corresponding energy-momentum tensor is
\begin{equation}
T^{\mu\nu}_{\rm em}=-F^{\mu\alpha}F^{\nu}_{\phantom{\nu}\alpha}
           +\squarter\eta^{\mu\nu}\,F_{\alpha\beta}F^{\alpha\beta}\,,
\label{eq:em-tensor-em}
\end{equation}
which we couple to the gravitational field according to
(\ref{eq:matter-coupling}). A priori, there is no reason why different
types of matter must couple with the same constant $\kappa$. However,
if we choose different constants for different matter types, say
$\kappa_1$ for type 1 and $\kappa_2$ for type 2, then the ratio
of gravitational to inertial mass of type 1 would differ from the
corresponding ratio of type 2 by a factor $\kappa_1/\kappa_2$ and hence
violate the weak equivalence principle, according to which the ratio
of gravitational to inertial mass is the same universal constant for all
types of matter. Hence, in order to conform with this principle, we
choose the same $\kappa$ for point particles and electromagnetism.
This is also the reason why we excluded a $h T$- coupling in
(\ref{eq:matter-coupling}).

Now we can play the same trick as for the point particle and
write~\cite{Th}
\begin{eqnarray}
S_{\rm em}+S^{\rm int}_{\rm em}
&=& - \squarter\int d^4x\
      \eta^{\alpha\mu}\eta^{\beta\nu}F_{\alpha\beta}F_{\mu\nu}
    - \tfrac{\kappa}{2}\int d^4x\ h_{\mu\nu} T^{\mu\nu}
\nonumber
\\
&=&
 - \squarter\int d^4x\ \sqrt{-\det\{g\}}\,
       g^{\alpha\mu} g^{\beta\nu}\,
   F_{\alpha\beta}F_{\mu\nu} + {\cal O}(\kappa^2)\,,
\label{eq:total-action-em}
\end{eqnarray}
where $g^{\mu\nu}$ and $\det\{g\}$ are the inverse and determinant of
the matrix $g_{\mu\nu}$. To first order in $\kappa$ we have
$g^{\mu\nu}=\eta^{\mu\nu}-\kappa h^{\mu\nu}$ and
$\sqrt{-\det\{g\}}=1+\tfrac{\kappa}{2} h$,
where $h=h^{\mu}_{\mu}$ and indices on $h$ are always raised and lowered
with $\eta$.

We are now in the same situation as for the point particle,
since (\ref{eq:total-action-em}) is the action for the electromagnetic
field on a spacetime with background metric $g$. Without sources its
equations of motion are
\begin{equation}
\partial _{[\mu} F_{\nu\lambda]}=0\quad
\partial_{\mu}\left(\sqrt{\det\{g\}}\,
g^{\mu\alpha}g^{\nu\beta}\,F_{\alpha\beta}\right)=0\,,
\label{eq:full-motion-em}
\end{equation}
which, in an eikonal approximation, imply that light rays are
lightlike geodesics in the metric $g$. Note that our decision
to couple all types of matter with the \emph{universal} coupling
constants has the effect that light rays and particle paths are
geodesics in the \emph{same} metric.

%---------------------------------------------------------------

\section{Deflection of light}
In this section we calculate the deflection of a light ray by
the spherically symmetric gravitational field (\ref{eq:solution1}).
We are interested in the amount of deflection as $m$ tends to zero.
A priori this limit is a precarious one since the metric coefficients
diverge as $m\rightarrow 0$ (see (\ref{eq:solution1})). However, we
now show that the deflection has a finite limit. To this end we split
(\ref{eq:solution1}) in a finite part,
\begin{equation}
h^{\rm f}_{\mu\nu}(\vec x) = -b\,
\left(
\begin{array}{cc}
f(r) & {\vec 0}^{\trans}\\
\vec 0 & \shalf\delta_{ij}f(r)
\end{array}
\right)\,,
\label{eq:finite-part}
\end{equation}
which has a continuous limit as $m\rightarrow 0$, and a diverging part,
\begin{equation}
h^{\rm d}_{ij}(\vec x) = bm^{-2}\,\partial_i\partial_j f(r)\,.
\label{eq:diverging-part}
\end{equation}
Since we strictly stay within the linear theory and, accordingly,
keep only terms linear in $\kappa$, the contributions of $h^{\rm f}$
and $h^{\rm d}$ to the deflection add linearly, so that we can consider
them separately. Expanding $f(r)=\exp(-mr)/r$ in powers of $m$, we
explicitly checked that no power smaller than 2 of $m$ in
(\ref{eq:diverging-part}) contributes to the deflection, so that
in the limit as $m\rightarrow 0$ the diverging part does not contribute
at all. In fact, one can argue that for each $m$ the whole diverging
part does not contribute to light deflection. The argument is as follows:
Recall that under a spatial coordinate transformation
$x^i\mapsto {x'}^i=x^i+\kappa k^i(x^j)$ the spatial components of the
metric $g_{ij}$ change -- to first order in $\kappa$ -- according to
$g_{ij}\mapsto g_{ij}-\kappa (k_{i,j}+k_{j,i})$. Hence, choosing
$k_i(x^j)=\shalf bm^{-2}\partial_if(r)$, we can remove $h^{\rm d}$
from $g=\eta+\kappa(h^{\rm f}+h^{\rm d})$. The resulting coefficients
$g_{\mu\nu}=\eta_{\mu\nu}+\kappa h^{\rm f}_{\mu\nu}$ are
then understood with respect to the coordinate system ${x'}^i$
(we drop the dash hereafter). Since $k^i$ falls of as
$r\rightarrow \infty$, the new coordinate axes asymptotically
approach the old ones. In particular, they are again asymptotically
orthogonal. Hence we may calculate the light deflection $\delta(m)$
as usual, using the metric $g^{\rm f}=\eta+\kappa h^{\rm f}$ with
the identification (\ref{eq:kappa-b-G-M}).
$\delta(m)$ is a continuous function of $m$ through the continuous
dependence of the light deflection on the background metric.
Hence, in order to obtain $\delta(0)$, we may calculate the light
deflection for the limit metric
\begin{equation}
\lim_{m\to 0}g^{{\rm f}}_{\mu\nu}(\vec x) =
\left(
\begin{array}{cc}
1 - \frac{2GM}{r} & {\vec 0}^{\trans}\\
\vec 0 & -\delta_{ij} \left( 1 + \frac{GM}{r} \right)
\end{array}
\right)\,.
\label{eq:solution-truncated}
\end{equation}
But instead of performing an explicit calculation,
we merely need to compare (\ref{eq:solution-truncated}) with the
Schwarzschild metric in General Relativity. In isotropic coordinates
the latter reads
\begin{eqnarray}
g^{\rm Schw}(\vec x)
&=&
\left(
\begin{array}{cc}
\left[
\frac{1-\frac{GM}{2r}}{1+\frac{GM}{2r}}\right]^2
&{\vec 0}^{\trans}\\
&\\
{\vec 0}
& -\delta_{ij}\,\left[1+\frac{GM}{2r}\right]^4
\end{array}
\right)\\
&=&
\left(
\begin{array}{cc}
1-\frac{2GM}{r}
& {\vec 0}^{\trans}
\\
&\\
{\vec 0}
& -\delta_{ij}\,\left(1+\frac{2GM}{r}\right)
\end{array}
\right) + {\cal O}(G^2) .
\label{eq:schwarzschild}
\end{eqnarray}
As is well known, in leading (linear) order of $G$, light deflection
receives equal contributions from the time-time and space-space parts:
\begin{equation}
\delta^{\rm Einst} = \frac{2GM}{q} + \frac{2GM}{q} = \frac{4GM}{q}\,,
\label{eq:einstein-bending}
\end{equation}
where $q$ is the impact parameter. Applied to our metric this
finally leads to the van Dam--Veltman value for the light deflection
in the zero-mass limit:
\begin{equation}
\lim_{m\to 0}\delta(m) = \frac{2GM}{q} + \frac{GM}{q} =
\frac{3GM}{q} = \frac{3}{4}\,\delta^{\rm Einst}\,.
\label{eq:vdv-bending}
\end{equation}

\vspace{1.0truecm}
\noindent
\textbf{Acknowledgements:} One of us (D.G.) likes to thank Norbert
Straumann for discussions, in the course of which the issue of a
classical understanding of the van~Dam--Veltman--discontinuity came up.

%------------------------------------------------------------------------

\end{document}